\documentstyle{article}
\bigskip
\begin{document}
\title{{\bf Classical and Quantum-like approaches
to Charged-Particle Fluids in a Quadrupole}}
\author{{\bf S. De Nicola}$^1$, {\bf R. Fedele}$^2$, and {\bf V.I.
Man'ko}$^{2,3}$ \\
{\it $^1$ Istituto di Cibernetica del CNR, Arco Felice, Italy}\\
{\it $^2$ Dipartimento di Scienze Fisiche, Universit\'{a} ``Federico II''}\\
{\it and INFN, Napoli, Italy}\\
{\it $^3$ P.N. Lebedev Physical Institute, Moscow, Russia}}
\maketitle
\begin{abstract}
A classical description of the dynamics of a dissipative charged-particle
fluid in a quadrupole-like device is developed. It is shown that the set of
the classical fluid equations contains the same information as a complex
function satisfying a Schr\"{o}dinger-like equation in which Planck's constant
is replaced by the time-varying emittance, which is related to the
time-varying temperature of the fluid. The squared modulus and the gradient
of the phase of this complex function are proportional to the fluid density
and to the current velocity, respectively. Within this framework, the
dynamics of an electron bunch in a storage ring in the presence of
radiation damping and quantum-excitation is recovered. Furthermore, both
standard and generalized (including dissipation) coherent states that may
be associated with the classical particle fluids are fully described in
terms of the above formalism.
\end{abstract}

\bigskip

PACS Number(s): {\it 52.65.Kj, 03.65, 29.20Dh, 41.85Lc}

\bigskip

Keywords: {\it Fluid equations, quantum-like descriptions,
quadrupoles, coherent states, storage rings}

\newpage
\section{Introduction}
The dynamics of a charged--particle system in a quadrupole-like
device concerns with a number of problems in accelerator physics
\cite{1,1bis}, in plasma physics \cite{2}, and in the physics of
electromagnetic (e.m.) traps \cite{3}. In particular,
charged--particles execute synchrotron oscillations in
radio--frequency (RF) fields; then, the system can be suitably
bunched as a result of the bucket formation \cite{4}. When
oscillations are sufficiently small, the syncrotron motion can be
described as being in a harmonic-like potential well whose strength is
proportional to the RF voltage amplitude \cite{4}. Typically, this
strength is a slow function of time compared to the RF period.

Sometimes, during the above dynamics in a RF field, the phenomenon
of dissipation due to the e.m. radiation emission cannot be
neglected. However, it allows the system to reach a steady state.

In the combined traps for positrons (electrons) and antiprotons
(protons), such a kind of dissipation enables the system to cool
and condense, and consequently to make possible the anti-hydrogen
(hydrogen) atom formation \cite{5}.

In the circular accelerating machines, the radiation damping, in
competition with the quantum excitation (photon noise), leads the
system to an asymptotic equilibrium state, which typically
corresponds to a Gaussian particle distribution in both
configuration space and momentum space, with a given final
temperature (or final emittance) \cite{1}. This final steady state
can be considered as a sort of {\it coherent state} reached by the
system which does not depend on its history \cite{1}. In fact, it
can be determined {\it a priori} in terms of both the bunch and
the machine parameters only \cite{1}. It is worth to note that
this phenomenon of dissipation does not always lead the bunch to
cool. In fact, the combined effect of radiation damping and
quantum excitation can make possible the warming up of the bunch.
It happens when the initial temperature (initial emittance) of the
bunch is smaller than the final one.

In this paper, we develop a fluid description for a charged-particle system
in a quadrupole-like potential well in the presence of dissipations in both
classical-like and quantum-like domains. In particular, a generalized class
of coherent states are shown to be possible, within this fluid framework,
in the presence of dissipation. It is shown that coherent states described
in the classical domain are equivalent to the ones described in the
quantum-like domain.

\section{Hydrodynamical description of coherent states for
charged-particle beams} In this section, following a recent hydrodynamic
approach applied to the {\it radiation fluids} \cite{6bis}, we start to
consider a classical-like description of coherent states for a dilute
charged-particle fluid, for which the ideal-gas-state equation is assumed.
For the sake of simplicity, we consider the 1-D motion fluid motion
equation of a dilute charged-particle beam. Accordingly, we have
\begin{equation}
\left( {\frac{\partial }{\partial s}} + P {\frac{\partial }{\partial x
}}%
\right) P = - {\frac{\partial U }{\partial x}} - {\frac{1 }{n}} {\frac{%
\partial \Pi }{\partial x}}~~~,  \label{a12}
\end{equation}
\begin{equation}
{\frac{\partial n }{\partial s}} + {\frac{\partial }{\partial x }}%
\left(nP\right) =0~~~,  \label{a13}
\end{equation}
where $s=ct$ ($c$, being the speed of light), $P=P(x,s)$ is the current
velocity, $n=n(x,s)$ is the particle number density, and the quantity $%
U=U(x,s)$ is a dimensionless effective potential acting on the system.
Assuming the ideal gas state equation
\begin{equation}
\Pi~=~ {\frac{nk_{\rm B} T }{mc^2}}~~~,  \label{ideal-gas-state}
\end{equation}
where $\Pi$ is the thermal pressure, normalized with respect to
$mc^2$, $k_{\rm B}$ is the Boltzmann constant, $m$ is the particle mass,
and $T =T(s) $ is the temperature of the system, we have
\begin{equation}
{\partial\Pi\over\partial x}~=~{\partial\Pi\over\partial n}{\partial
n\over\partial x}~=~\eta {\partial n\over\partial x}~~~,
\label{eta-definition}
\end{equation}
with the definition
\begin{equation}
\eta \equiv {\frac{\partial \Pi }{\partial n}}~~~.
\label{eta-def}
\end{equation}
Note that, for an isothermal transformation,
\begin{equation}
\eta ~=~{\frac{k_{\rm B} T }{mc^2}}~=~{v_{\rm th}^{2}\over c^2}~
=~\mbox{const}.~~~,
\label{eta-isothermal}
\end{equation}
where $v_{\rm th}\equiv \sqrt{k_{\rm B} T/m}$
is the thermal velocity. In
general, the explicit expression of $\eta $ depends on the
particular thermodynamical transformation that the system
undergoes, but it is related someway to the r.m.s. of
momentum-space distribution $\sigma_P$.
Let us assume, in the following, that $P$ and $\eta$ are functions of
$s$ only, namely,
\begin{equation}
P(x,s)~=~ P_{0}(s)  \label{P-zero}
\end{equation}
and
\begin{equation}
\eta(x,s)~=~ \eta_{0}(s)~~~.  \label{eta-zero}
\end{equation}
Consequently, (\ref{a12}) and (\ref{a13}) become
\begin{equation}
P^{^{\prime}}_0 (s)~=~ -{\frac{\partial U}{\partial x }}- \eta_0
(s){\frac{%
\partial}{\partial x}}\ln n~~~,  \label{motion-eq}
\end{equation}
\begin{equation}
{\frac{\partial n}{\partial s}}~+~P_0 (s) {\frac{\partial n}{\partial
x}}%
~=~0~~~,  \label{continuity-eq}
\end{equation}
where the prime denotes differentiation with respect to $s$.
Thus, (\ref{motion-eq}) can be easily integrated with respect to $x$,
giving
\begin{equation}
n(x,s)~=~\exp\left\{-{\frac{1}{\eta_0 (s) }}\left[U(x,s)+P^{^{%
\prime}}_{0}(s)x+g(s)\right] \right\}~~~,  \label{n-solution}
\end{equation}
where $g(s)$ is an arbitrary function of $s$. By substituting (\ref
{n-solution}) in (\ref{continuity-eq}), we obtain
\begin{equation}
{\frac{\eta^{^{\prime}}_0
}{\eta_0}}\left(U+P^{^{\prime}}_{0}x+g\right)-%
\left({\frac{\partial U}{\partial s }}+P^{^{\prime\prime}}_0 x +
g^{^{\prime}}\right)-P_0 \left({\frac{\partial U}{\partial x }}%
+P^{^{\prime}}_{0}\right)=0~~~.  \label{continuity-1}
\end{equation}
In order to consider the special case of coherent state associated with
the
beam, let us assume that the potential $U(x,s)$ is given by
\begin{equation}
U(x,s)= {\frac{1}{2}}K(s)x^2~~~.  \label{quadrupole}
\end{equation}
Equation (\ref{quadrupole}) defines a quadrupole-like potential
well with time-varying strength $K(s)$. Substituting
(\ref{quadrupole}) in
(\ref{continuity-1}%
) we get the following conditions for the quantities $\eta_0$, $P_0$,
$K$, and $g$:
\begin{equation}
{\frac{\eta_0^{^{\prime}} }{\eta_0}}={\frac{K^{^{\prime}}}{K}}~~~,
\label{eta-K}
\end{equation}
\begin{equation}
P^{^{\prime\prime}}_0 -{\frac{\eta_0^{^{\prime}} }{\eta_0}}P^{^{%
\prime}}_0+KP_0=0~~~,  \label{P-zero-eq}
\end{equation}
\begin{equation}
{\frac{d}{ds}}\left[{\frac{1}{2}}P^{2}_0 +
g\right]={\frac{\eta_0^{^{\prime}}
}{\eta_0}}g~~~.  \label{energy-eq}
\end{equation}
>From (\ref{eta-K}), we obtain
\begin{equation}
\beta\equiv {\frac{\eta_0 (s)}{K(s)}}=\mbox{const}.  \label{alpha-def}
\end{equation}
On the other hand, substituting (\ref{quadrupole}) and (\ref{alpha-def})
into (\ref{n-solution}), we have
\begin{equation}
n(x,s)~=~\exp\left\{-{\frac{\left[x-x_0 (s)\right]^{2}
}{2\beta}}\right\}~~~,
\label{n-solution-bis}
\end{equation}
where
\begin{equation}
x_0 (s)\equiv -{\frac{2g(s)}{P^{^{\prime}}_0 (s)}}~~~,  \label{x-zero}
\end{equation}
and
\begin{equation}
g(s)= {\frac{1}{2}}{\frac{\eta_0 (s)}{\beta}}x^{2}_{0}(s)={\frac{1}{2}}%
K(s)x^{2}_{0}(s)~~~.  \label{g-eq}
\end{equation}
By combining (\ref{x-zero}) and (\ref{g-eq}), we find
\begin{equation}
P^{^{\prime}}_{0}(s)=-K(s)x_0 (s)~~~.  \label{g-eq-bis}
\end{equation}
Furthermore, substituting (\ref{n-solution-bis}) in
(\ref{continuity-eq}),
we obtain
\begin{equation}
P_{0}(s)=x^{^{\prime}}_{0} (s)~~~.  \label{linear-momentum}
\end{equation}
Combining the results given by
(\ref{eta-K})--(\ref{linear-momentum}), we
finally obtain
\begin{equation}
x^{^{\prime\prime}}_{0}+K(s)x_{0}=0~~~,  \label{x-zero-eq}
\end{equation}
\begin{equation}
P^{^{\prime\prime}}_{0}-\Gamma (s)P^{^{\prime}}_{0}+K(s)P_{0}=0~~~,
\label{P-zero-eq-bis}
\end{equation}
and
\begin{equation}
{\frac{d}{ds}}\left[{\frac{1}{2}}P^{2}_0 +
{\frac{1}{2}}Kx^{2}_{0}\right]=%
\Gamma (s)\left({\frac{1}{2}}Kx^{2}_{0}\right)~~~,
\label{energy-eq-bis}
\end{equation}
where
\begin{equation}
\Gamma (s)\equiv {\frac{\eta_0^{^{\prime}}
}{\eta_0}}={\frac{K^{^{\prime}}}{K}}%
~~~.  \label{Gamma-def}
\end{equation}
In conclusion, a normalized density solution of the system of Eqs.~(\ref
{motion-eq}) and (\ref{continuity-eq}) associated with potential (\ref
{quadrupole}) is
\begin{equation}
n(x,s)~=~{\frac{1}{\sqrt{2\pi\beta}}}\exp\left\{-{\frac{\left[x-x_0
(s)\right]^{2} }{2\beta}}\right\}~~~.  \label{n-solution-ter}
\end{equation}
>From (\ref{n-solution-ter}), it is clear that the beam size
(r.m.s.) $\sigma_0$ is
\begin{equation}
\sigma_0 \equiv \langle\left(x-x_0\right)^2\rangle^{1/2}=\sqrt{\beta}%
=\mbox{const}.~~~,  \label{sigma-zero}
\end{equation}
which implies that $\beta$ must be positive. It is clear that the
Gaussian solution (\ref{n-solution-ter}) with (\ref{x-zero-eq})
and the positivity of $\beta$ may be a coherent state associated
with the charged-particle beam in a quadrupole-like potential well
with also a time-varying strength $K(s)$. But, in this case,
(\ref{P-zero-eq-bis})--(\ref{Gamma-def}) show that this coherent
state exists for a dissipative system, otherwise $K$ must be
constant. In fact, the $s$-variation of $\eta_0$ implies that the
system, in principle, is not closed and can exchange its energy
with the environment (see
Eqs.~(\ref{x-zero-eq})--(\ref{energy-eq-bis})).
This way, the distribution of
the particles around the center remains unchanged (see
Eqs.~(\ref{n-solution-ter}) and (\ref {sigma-zero})) and corresponds to
the coherent-beam configuration, for which the following condition
holds:
\begin{equation}
\eta_0 (s) = \sigma^{2}_{0}K(s)~~~.  \label{matching-condition}
\end{equation}
Let us now consider, as a special case, an isothermal
transformation of the beam, for which $\eta_0$ is constant. This
is equivalent to say that the beam emittance $\epsilon$ \cite{7}
is constant too
\begin{equation}
{\frac{\epsilon^2}{4\sigma^{2}_{0}}} =
{\frac{v^{2}_{\rm th}}{c^2}}=\mbox{const}.
\label{emittance}
\end{equation}
According to (\ref{eta-K}) and (\ref{P-zero-eq}), it is clear that
a coherent state exists, in this case, if and only if, $K$ is
constant. Moreover, (\ref{x-zero-eq})--(\ref{energy-eq-bis}) become
\begin{equation}
x^{^{\prime\prime}}_{0}+K x_{0}=0~~~,  \label{x-zero-eq-bis}
\end{equation}
\begin{equation}
P^{^{\prime\prime}}_{0}+K P_{0}=0~~~,  \label{P-zero-eq-ter}
\end{equation}
and
\begin{equation}
{\frac{1}{2}}P^{2}_0 + {\frac{1}{2}}K x^{2}_{0}\equiv {\cal
E}_0=\mbox{const}.
\label{energy-eq-ter}
\end{equation}
Additionally, (\ref{matching-condition}) becomes
\begin{equation}
K\sigma^{4}_{0}={\frac{\epsilon^2}{4}}~~~.
\label{matching-condition-bis}
\end{equation}
In this case, (\ref{n-solution-ter}) with
(\ref{x-zero-eq-bis})--(\ref {matching-condition-bis}) represent
a coherent structure which preserves both the energy and the beam
emittance. We may refer to this kind of coherent states as {\it
ordinary} or {\it isothermal coherent states}.
Furthermore, (\ref{Gamma-def}) can be written as
\begin{equation}
\Gamma (s)={\frac{1}{\epsilon ^{2}}}{\frac{d\epsilon ^{2}}{ds}}~~~.
\label{emittance-eq}
\end{equation}
Let us now observe that, on the basis of (\ref{linear-momentum}),
(\ref{x-zero-eq}), (\ref{n-solution-ter}), and (\ref{sigma-zero}), one
can introduce the following complex function
\begin{equation}
\Psi (x,s)~=~\sqrt{n(x,s)}\exp {i\theta (x,s)}~~~,
\label{complex-function}
\end{equation}
where $\theta (x,s)$ is defined by
\begin{equation}
P_0 (s)~=~\epsilon (s) {\partial\theta (x,s)\over\partial x}~~~.
\label{theta-def}
\end{equation}
Consequently, (\ref{complex-function}) can be explicitly written as
\begin{equation}
\Psi (x,s)~=~\frac{1}{\left(2\pi \sigma_{0}^{2} \right)^{1/4}} \exp
\left[ -{\frac{\left[x-x_{0}(s)\right]^{2}}{{4\sigma_{0} ^{2}}}}
+{\frac{i P_0 (s)x}{{2\epsilon (s)}}}+i\phi_0 (s)\right] ~~~,
\label{explicit-psi}
\end{equation}
where $\phi_0 (s)$ is an arbitrary function of $s$. It is worth to
note that the above complex function $\Psi$ contains the same
information as the system (\ref{linear-momentum}),
(\ref{x-zero-eq}), (\ref{n-solution-ter}), and (\ref{sigma-zero}).
However, Eq. (\ref{explicit-psi}) recovers a coherent state
(similar to the ones in the ordinary quantum mechanics \cite{8}),
when $\epsilon$ is assumed to be constant (non-dissipative
system).

The constant $\epsilon$ plays the similar role as $\hbar$ in quantum
mechanics. Consequently, in this case, the complex function defined by Eq.
(\ref{explicit-psi}) satisfies a Schr\"{o}dinger-like equation for a harmonic
oscillator potential $U=Kx^2 /2$, where $K$ is assumed to be constant, as
well. We point out that the above isothermal coherent states coincide with
the coherent states for charged-particle beam that have been recently
described, within a wave-like context, in Ref.s \cite{6}.

Additionally, when $\epsilon$ and $K$ depend both on $s$, it has
already been proven that Eq.~(\ref{explicit-psi}) still describes
a quantum-like coherent state, provided that Eq. (\ref{emittance})
is satisfied \cite{9,10}. The coherent states of a driven
oscillator with dissipation within the framework of ordinary
quantum mechanics were constructed in~\cite{DodMan79}.

\section{Quantum-like description}
On the basis of the results given in the previous section, we extend the
above correspondence between the classical and the quantum-like description
\cite{fs} of the fluid motion, beyond the coherent state assumption. To
this regard, we assume, in the absence of collective effects, that the
dynamics of our system is governed by the following Schr\"{o}dinger-like
equation:
\begin{equation}
i\alpha {\frac{\partial \psi }{\partial s}}~=~-{\frac{\alpha
^{2}}{2}}{\frac{\partial ^{2}\psi }{\partial x^{2}}}+U(x,s)\psi ~~~,
\label{schro}
\end{equation}
where $\alpha =\alpha (s)$ plays the role of a dispersion parameter and
$U$, $x$ and $s$ have the same meaning as in the previous section. We show
that the above Schro\"{o}dinger-like equation is equivalent, under certain
conditions, to the classical fluid system of equations (\ref{a12}) and
(\ref{a13}). In fact, if we write
\begin{equation}
\psi (x,s)~=~M(x,s)\exp {i\varphi (x,s)}~~~,  \label{defpsi}
\end{equation}
and if we substitute (\ref{defpsi}) back into (\ref{schro}) we can easily
derive the following dissipative Madelung-like fluid equations, namely,
\begin{equation}
\left( {\frac{\partial }{\partial s}}+P{\frac{\partial }{\partial
x}}\right)
P=-{\frac{\partial U}{\partial x}}+{\frac{\alpha ^{^{\prime }}}{\alpha
}}P+{\frac{\alpha ^{2}}{2}}{\frac{\partial }{\partial x}}\left(
{\frac{1}{M}}{\frac{\partial ^{2}M}{\partial x^{2}}}\right) ~~~,
\label{mad1}
\end{equation}
\begin{equation}
{\frac{\partial M^{2}}{\partial s}}+{\frac{\partial }{\partial x}}\left(
M^{2}P\right) =0~~~,  \label{mad2}
\end{equation}
where
\begin{equation}
P=\alpha {\frac{\partial \varphi }{\partial x}}~~~.  \label{current}
\end{equation}
Note that we can define the fluid density
\begin{equation}
n(x,s)={|\psi (x,s)|}^{2}=M^{2}(x,s)~~~.  \label{nmdefinition}
\end{equation}
Consequently, under the following condition
\begin{equation}
{\frac{\alpha ^{^{\prime }}}{\alpha }}P+{\frac{\eta_0
}{n}}{\frac{\partial n}{\partial x}}+{\frac{\alpha
^{2}}{2}}{\frac{\partial }{\partial x}}\left(
{\frac{1}{M}}{\frac{\partial ^{2}M}{\partial x^{2}}}\right) =0~~~,
\label{basiccondition}
\end{equation}
(\ref{mad1}) reduces to the following classical-like form
\begin{equation}
\left( {\frac{\partial }{\partial s}}+P{\frac{\partial }{\partial
x}}\right)
P=-{\frac{\partial U}{\partial x}}-{\frac{\eta_0 }{n}}{\frac{\partial
n}{\partial x}}~~~,  \label{reduced}
\end{equation}
where $\eta_0 =\eta_0 (s)$ has been already defined above. It is
clear that (\ref {mad2}) and (\ref{reduced}) together with the
quantum-like interpretation (\ref{nmdefinition}) formally concide
with our starting classical system as given by (\ref{a12}) and
(\ref{a13}).
Now we show that the classical-like solution for the dissipative
Schr\"{o}dinger-like equation (\ref{schro}) satisfying
(\ref{basiccondition}) can effectively be determined in the case
of a quadrupole-like potential, i.e., $U=K(s)x^2 /2$, where $K(s)$
is the quadrupole strength. Indeed, in this case, Eq.~(\ref{schro})
admits the following solution:
\begin{equation}
\psi ~=~\frac{1}{\left[2\pi \sigma^{2} (s)\right]^{1/4}}\exp \left[
-{\frac{x^{2}}{{4\sigma ^{2} (s)}}}+{\frac{ix^{2}}{{2\alpha (s)\rho
(s)}}}+i\chi (s)\right] ~~~.  \label{gaussian}
\end{equation}
>From (\ref{gaussian}) and (\ref{current}), we obtain the following
expression for the current velocity:
\begin{equation}
P(x,s)~=~-{\frac{x}{\rho (s)}}~~~.  \label{pgauss}
\end{equation}
In (\ref{gaussian}) and (\ref{pgauss}),
$\sigma (s)$, $\rho (s)$, and $\chi (s)$ are real functions satisfying
the
following set of differential equations:
\begin{equation}
{\frac{1}{\rho }}~=~{\frac{1}{\sigma }}{\frac{d\sigma }{ds}}~~~,
\label{1suro}
\end{equation}
\begin{equation}
{\frac{d\chi }{ds}}~=~-{\frac{\alpha (s)}{{4\sigma^{2} (s)}}}~~~,
\label{2suro}
\end{equation}
\begin{equation}
{\frac{d^{2}\sigma }{ds^{2}}}+K(s)\sigma -{\frac{1}{\alpha
}}{\frac{d\alpha
}{ds}}{\frac{d\sigma }{ds}}-{\frac{\alpha ^{2}}{{4\sigma ^{2}}}}~=~0~~~.
\label{3suro}
\end{equation}
Up to this point, the function $\alpha (s)$ is quite arbitrary within a
purely
quantum-like context. However, we point out that, by substituting (\ref
{gaussian}) into (\ref{basiccondition}), in view of
(\ref{defpsi}), (\ref{current}), and (\ref{nmdefinition}),
the previous equations (\ref{pgauss})--(\ref{3suro})
are exactly obtained, provided that the following condition for
$\alpha (s) $ is satisfied:
\begin{equation}
\eta_0 (s)={\frac{{\sigma }}{{\alpha }}\frac{d\alpha }{ds}\frac{d\sigma
}{ds}}+{\frac{\alpha ^{2}}{4\sigma ^{2}}}~~~.  \label{alfaeta}
\end{equation}
This condition clearly shows that $\alpha (s)$ is essentially
determined by the temperature $T(s)$ of the fluid through $\eta_0
(s)$ (see Eqs.~(\ref{ideal-gas-state}), (\ref{eta-def}), and
(\ref{eta-zero})~). On the other hand, within the quantum-like
framework, the r.m.s of the momentum distribution $\sigma _{P}$ is
defined as
\begin{equation}
\sigma _{P}(s)=\alpha \left[ \int_{-\infty }^{+\infty }\left|
{\frac{\partial \psi (x,s)}{\partial x}}\right| ^{2}~dx\right] ^{1/2}~=
\left[\left( ~{\frac{d\sigma }{ds}}\right) ^{2}+{\frac{\alpha
^{2}}{{4\sigma
^{2}}}}\right]^{1/2}~~~.  \label{sigmapi}
\end{equation}
Note that, consistently with the quantum-like formalism, the above
definition of the r.m.s. of the momentum distribution, suggests to make the
following assumption:
\begin{equation}
\eta_0 (s)={\sigma _{P}^{2}(s)}~~~.  \label{eta-sigma}
\end{equation}
On the other hand, in the classical-like interpretation, $\sigma
_{P}(s)$ is r.m.s. of a Maxwellian-like (Gaussian) distribution in
the momentum space and, consequently, it is proportional to the
square root of the temperature of the system (see
Eq.~(\ref{reduced})~). By inserting (\ref{alfaeta}) and (\ref{sigmapi})
into (\ref{eta-sigma}), we obtain the following condition:
\begin{equation}
{\frac{1}{\alpha }}{\frac{d\alpha }{ds}}={\frac{1}{\sigma
}}{\frac{d\sigma }{ds}}~~~.
\label{alfa-sigma}
\end{equation}

\section{Connection between $\alpha$ and the beam emittance
$\epsilon$} Within the classical framework, it is well known that
the beam emittance $\epsilon$ can be obtained by the relation
\cite{7}
\begin{equation}
{\epsilon^2\over 4} = \langle x^2 \rangle ~ \langle p^2 \rangle -
\langle xp \rangle^2~~~,  \label{xpi}
\end{equation}
where $\langle x^2 \rangle = \sigma^2 $, $\langle p^2\rangle =
\sigma_{P}^2$, and $\langle xp\rangle^2={\sigma^2}{\left( d\sigma
/ds \right)^2}$; the averages are taken with respect to the
classical phase-space Gaussian distribution whose configuration
projection coincides with ${{\vert \psi \vert}^2}$. Taking into
account the above relations,  (\ref{xpi}) can be written as
\begin{equation}
\sigma_P^2 = \left({\frac{d\sigma}{ds}}\right)^{2}+
{\frac{\epsilon^{2}}{{4\sigma^{2}}}}~~~.  \label{epssigma}
\end{equation}
Consequently, by comparing (\ref{epssigma}) and (\ref{sigmapi}), we
obtain
the following equality:
\begin{equation}
\epsilon \left( s\right) =\alpha \left( s\right) ~~~,
\label{epsialfa}
\end{equation}
and the envelope equation (\ref{3suro}) becomes
\begin{equation}
{\frac{d^2 \sigma}{ds^2}}+K(s)\sigma- \left({\frac{1}{\epsilon}}
{\frac{d\epsilon}{ds}}\right){\frac{d\sigma}{ds}}-
{\frac{\epsilon^2}{4\sigma^2}}~~~, \label{envelope-equation}
\end{equation}
which coincides with the envelope equation obtained in
Ref.~\cite{9}.

\section{Conclusions and remarks} We have proven that a dissipative
classical fluid, moving in a quadrupole-like focusing device, can
be fully described in terms of a Schr\"{o}dinger-like equation for
harmonic oscillator where the Planck's constant is replaced by the
time-varying beam emittance. This result justifies the main
assumption of Refs.~\cite{9,10} where the longitudinal dynamics of
an electron bunch in a storage ring in the presence of radiation
damping and quantum excitation has been described by
Eqs.~(\ref{schro}) and (\ref{epsialfa}). We point out that
coherent state, in the dissipative case, are recovered from Eq.
(\ref{envelope-equation}) in the limit of $1/\rho \rightarrow 0$,
namely, $d\sigma /ds =0$ for any $s$ ($\sigma =\sigma_0$ for any
$s$), to give Eq.~(\ref{matching-condition}).  Furthermore, in
this case, Eq.~(\ref{epssigma}) reduces to the minimum uncertainty
relation condition, which is a typical feature of coherent states.
Remarkably, note that, in the absence of dissipation ($\epsilon =
\mbox{const}$.), all the results of Ref. \cite{6}, concerning with
coherent states of charged-particle beams, are fully recovered by
the present treatment, provided that $K$ is assumed constant, as
well. Finally, we point out that the fluid treatment presented in
this paper can be also applied to the e.m. traps \cite{11} with
the inclusion of the dissipation.

\section*{Acknowledgments}
V.I.M. would like to acknowledge Dipartimento di Scienze Fisiche,
Universit\'{a} ``Federico II'' di Napoli for kind hospitality and
the Russian Foundation for Basic Research for the partial support
under Project~No.~99-02-17753.


\begin{thebibliography}{99}
\bibitem{1}M. Sands, ``The Physics of electron storage rings.
An Introduction,'' in: Proc. Int. School of Physics ``Enrico Fermi,''
Course XLVI, Varenna, 16-26 June 1969, B. Touschek (ed.), (Academic
Press, New York, 1971), p.~257.
\bibitem{1bis}H. Wiedemann, {\it
Particle Accelerator Physics} (Springer Verlag, Berlin, 1993).
\bibitem{2}F.F. Chen, {\it Introduction to Plasma Physics} (Plenum
Publishers, New York, 1974).
\bibitem{3}T.M. O'Neil and T.F. Driscol, {\it Phys. Fluids},
{\bf 22}, 266 (1979).
\bibitem{4}B.W. Montague, ``RF Acceleration,'' in: Proc. First
Course of the Int. School of Particle Acc. of the ``Ettore
Majorana'' Centre for Scientific Culture, Erice, 10-22 November,
1976, M.H. Blewett (ed.), CERN 77-13 (CERN, Geneva, 1977), p.~63.
\bibitem{5}D.J. Wineland, in: {\it Cooling, Condensation, and Storage
of Hydrogen Cluster Ions Workshop}, SRI International, Menlo Park,
California, 8-9 January, 1987; G.~Gabrielse, S.L.~Rolston,
L.~Haarmsa, and W.~Kell, Phys. Rev. A, {\bf 129}, 38 (1988).
\bibitem{6} S. De Nicola, R. Fedele, V.I. Man'ko, and G. Miele, {\it
Phys. Scr.}, {\bf 52}, 191 (1995).
\bibitem{6bis}R. Fedele, and V.I. Man'ko, {\it Phys. Rev. E} {\bf 60}
6042 (1999).
\bibitem{7}J. Lawson {\it The Physics of Charged-Particle Beams}
(Clarendon Press, Oxford, 1988), 2nd edition.
\bibitem{8}L.D. Landau and E.M. Lifshitz,
{\it Quantum Mechanics} (Pergamon Press, Oxford, 1965).
\bibitem{9}R. Fedele, G. Miele and L. Palumbo, {\it Phys. Lett. A},
{\bf 194}, 113 (1994).
\bibitem{10}S. De Martino, S. De Siena, and R. Fedele, {\it Phys.
Scr.} {\bf 56}, 426 (1997).
\bibitem{DodMan79}V.V. Dodonov and V.I. Man'ko, {\it Phys. Rev.}
A, {\bf 20}, 550 (1979).
\bibitem{fs}R. Fedele and P.K. Shukla (ed.s), {\it Quantum-like Model
and Coherent Effects} (World Scientific, Singapore, 1995).
\bibitem{11}R. Fedele, R. Poggiani, and G. Torelli, ``A thermal wave
model for electromagnetic traps,'' in: Proc. of INFN Elosatron
Project 31st Workshop ``Cristalline Beams and Related Issues,''
Erice, Italy, 11-21 November 1995, A.G.~Ruggiero (ed.), (1996),
p.~393; R.~Fedele, G.~Gorini, G.~Torelli, and D.~Zanello,
``Quantum-like description of charged-particle traps''  (to be
published in Proc. Sixth Int. Conf. on Squeezed States and
Uncertainty Relations, Napoli, Italy, 24-29 May 1999); R.~Fedele,
G.~Gorini, G.~Torelli, and D.~Zanello, ``Quantum-like equilibrium
states in a nested trap'' (submitted to {\it Phys. Scr.}, 1999).
\end{thebibliography}
\end{document}